\documentclass[pre,showpacs]{revtex4}
\usepackage{amssymb}
\usepackage[dvips]{graphicx}
\usepackage{amsmath}
\usepackage{graphicx}
\usepackage{makeidx}
\usepackage{subfigure}
\begin{document}
\title{Cooperative jump motions of jammed particles in a one-dimensional periodic potential}
\author{Hidetsugu Sakaguchi}
\affiliation{Department of Applied Science for Electronics and Materials,
Interdisciplinary Graduate School of Engineering Sciences, Kyushu
University, Kasuga, Fukuoka 816-8580, Japan}

\begin{abstract}
Cooperative jump motions are studied for mutually interacting particles in a one-dimensional periodic potential. The diffusion constant for the cooperative motion  in systems including a small number of particles is numerically calculated and it is compared with theoretical estimates. We find that the size distribution of the cooperative jump motions obeys an exponential law in a large system. 
\end{abstract}
\maketitle

Glass transitions and jamming transitions have been intensively studied for supercooled liquids and granular materials~\cite{rf:1}. 
The dynamics becomes very slow near the glass and jamming  transitions. Furthermore, the dynamical heterogeneity is observed, in which some regions exhibit faster dynamics than the rest region~\cite{rf:2,rf:3,rf:4}. In the mobile regions, cooperative jump motions and string-like cooperative motions are often observed in numerical simulations of supercooled liquid~\cite{rf:5,rf:6}. Donati et al. found that the probability distribution of the string length obeys an exponential law~\cite{rf:6}. However, the mechanism of the cooperative jump motions is not completely understood in the supercooled liquids.  We tried to understand the cooperative jump motion from a viewpoint of dynamical systems~\cite{rf:7}. There is a delocalization transition in a system of a small number of particles confined in a box with periodic boundary conditions. It is  a kind of chaos-chaos transitions in chaotic dynamical systems. Near the delocalization transition, the particles exhibit cooperative jump motions.
 
Near the glass transitions of supercooled liquids, the particles tend to be trapped in cages, which are constructed of the jammed particles by themselves.  In this brief report, we propose a one-dimensional system of $N$ particles under a periodic potential as shown in Fig.~1(a) as one of the simplest models which exhibit cooperative jump motions. The external periodic potential works as a cage for each particle in this model. We study a mechanism and statistical properties of the cooperative jump motions in this simple model.  

The model equation has a form of an overdamped Langevin equation: 
\begin{equation}
\frac{dx_i}{dt}=-F_0\sin(2\pi x_i)+F(x_i-x_{i-1})-F(x_{i+1}-x_{i})+\xi_i(t),
\end{equation}
where $F_0$ denotes the amplitude of the spatially periodic force, and the noise $\xi_i(t)$ satisfies $\langle \xi_i(t)\xi_j(t^{\prime})\rangle=2T\delta_{i,j}\delta(t-t^{\prime})$.  The repulsive interaction between the neighboring particles is expressed as $F(x)=-\partial U/\partial x$, using the Lennard-Jones potential $U(x)=-\epsilon(\sigma^6/x^6-\sigma^{12}/x^{12})$ for $x<x_c=2^{1/6}\sigma$. The attractive part of the Lennard-Jones potential is neglected for the sake of simplicity, i.e., $F(x)=0$ for $x>x_c=2^{1/6}\sigma$.  $N$ particles are confined in a system of size $N$ and the periodic boundary conditions are imposed. 
That is, $x_i$ is reset to $x_i-N$ when $x_i$ reaches $N$, and $x_i$ is reset to $x_i+N$ when $x_i$ reaches $0$. If $x_c>1$, the particles are jammed by the repulsive interaction. The spatially periodic force is derived from the external potential $U_0(x)=-F_0/(2\pi)\cos(2\pi x)$. 
In the state of the lowest energy, each particle is located at the minimum point $x=i$ of the external potential. 
 If the thermal noise is sufficiently weak, the particles are confined around the potential minima. If the noise strength becomes larger, the particles overcome the potential peaks and exhibit a random walk, although there is no definite phase transition in this one-dimensional system. 

We have performed numerical simulations at $\epsilon=0.01$ and $F_0=0.8$. The initial positions are $x_i(0)=i$.  
Figure 1(b) displays time evolutions of $\tilde{x}_i(t)$ for $i=1,3,5$, and 7 for $T=0.18$, $N=8$, and $\sigma=1/2^{1/6}$. 
Here, $\tilde{x}_i(t)$ is the position of the $i$th particle neglecting the resetting process $x_i(t)\rightarrow x_i(t)\pm N$ at the two boundaries $x=0$ and $x=N$.  Each particle is fluctuating near the potential minima $x=i$ for most of the time, but occasionally jumps to the neighboring potential minimum $i\pm 1$. All particles jumps almost simultaneously as shown in Fig.~1(b). 
Figure 1(c) shows time evolutions of $x_5(t)$, $x_2(t)-x_1(t)$, and $\Delta x=\sum_{j=1}^N(x_{j+1}-x_{j}-1)^2$.  The difference $x_2-x_1$ changes in time randomly. That is, individual motions are not synchronized, although the jump motions are well synchronized. 
The total fluctuation expressed by $\Delta x$ tends to increase when the jump motions occur. 

Each particle exhibits a random motion around potential minima almost independently. One particle might try to jump to the neighboring potential valley by the random motion, but there is another particle around the neighboring potential minimum, and two particles are hardly confined in the same potential valley, because of the strong repulsive interaction. As a result, the jump motions are permitted, only when all particles jump almost simultaneously to the right or to the left. Therefore, the probability of the simultaneous jump motion becomes rapidly smaller, as the total number $N$ is increased for a fixed value of $T$.

\begin{figure}[tbp]
\begin{center}
\includegraphics[height=8.5cm]{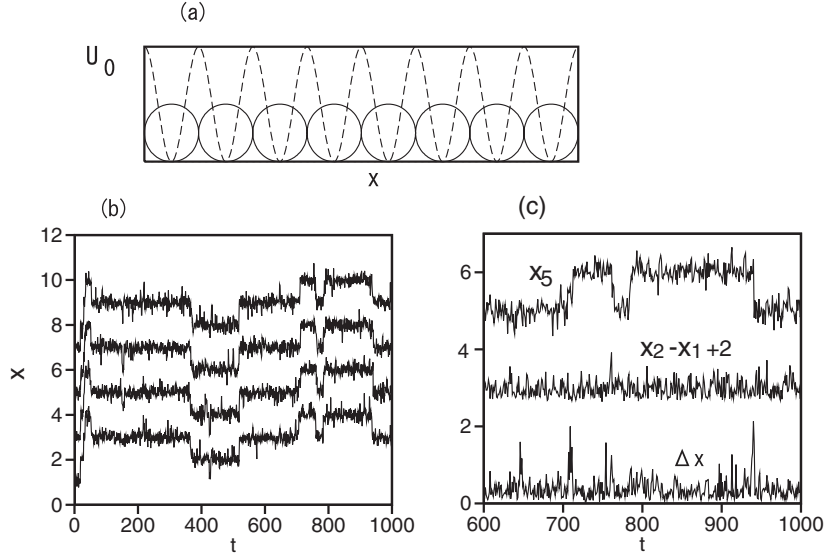}
\end{center}
\caption{(a) Schematic figure of mutually interacting particles in a spatially periodic potential.  (b) Time evolution of $x_i(t)$ for $i=1,3,5$ and 7. Cooperative jump motions are clearly observed. (c) Time evolutions of $x_5(t)$, $x_2(t)-x_1(t)$ (shifted by 2 for the sake of visibility), and $\Delta x=\sum_{j=1}^N(x_{j+1}-x_{j}-1)^2$ for $600<t<1000$.}
\label{f1}
\end{figure}

The jump probability can be evaluated with the diffusion constant. The diffusion constant is numerically calculated with $D=\langle (x_i(t^{\prime})-x_i(t))^2\rangle/\{2(t^{\prime}-t)\}$ for a sufficiently large interval $t^{\prime}-t$. 
The larger $T$ is necessary for the particles to overcome the potential peaks, when the particle number $N$ is increased. The diffusion constant decreases rapidly with $N$.  We shows the diffusion constant $D$ as a function of $T/N$ in Fig.~2(a) at $N=1,2,4$ and 8 for $x_c=1$. In our numerical simulation, $\langle \cdots\rangle$ was calculated as a simple average of 80,000 samples and $t^{\prime}-t$ was set to be $1000$ to evaluate $D$.  The diffusion constant $D$ is roughly approximated as a function of $T/N$. For $N=1$, the diffusion constant is exactly expressed as 
\begin{equation}
D(T)=T\left\langle \exp(U_0/T)\right\rangle^{-1}\left\langle \exp(-U_0/T)\right\rangle^{-1},
\end{equation}
where $\langle \exp(\pm U_0/T)\rangle=\int_0^{1}\exp(\pm U_0(x)/T)dx$~\cite{rf:8,rf:9,rf:10}. 
The numerical results for $N=1$ are well approximated at the theoretical curve.  If the motions are assumed to be sufficiently synchronized, the center of mass $X=(1/N)\sum_{i=1}^N \tilde{x}_i$ obeys
\begin{figure}[tbp]
\begin{center}
\includegraphics[height=5.cm]{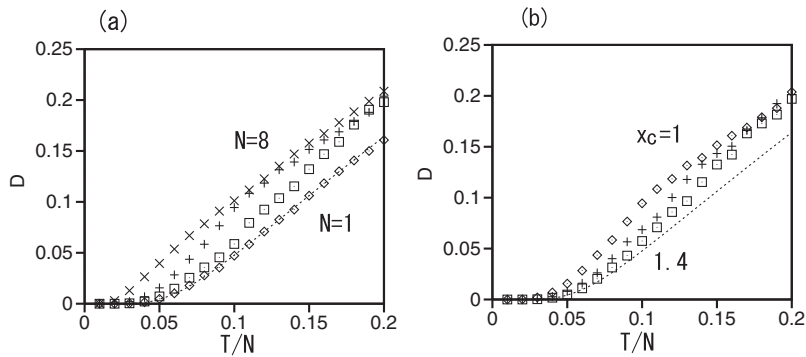}
\end{center}
\caption{(a) Diffusion constant $D$ as a function of $T/N$ for $N=1$ ($\Diamond$) ,2 ($\Box$) ,4 ($+$) and 8 ($\times$). The parameter value of $x_c$ is 1. The dotted curve denotes Eq.~(2). (b) Diffusion constant $D$ as a function of $T/N$ for $N=4$. The parameter $x_c$ are changed as $x_c=1$ ($\Diamond$), 1.2 ($+$)  and 1.4 ($\Box$). The dotted curve denotes Eq.~(2).}
\label{f2}
\end{figure}
\begin{figure}[tbp]
\begin{center}
\includegraphics[height=5.cm]{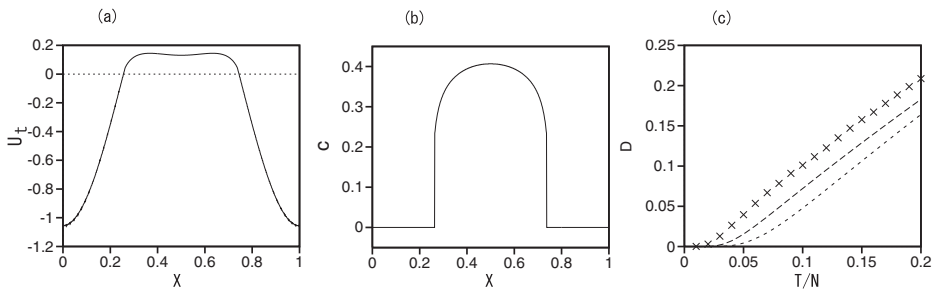}
\end{center}
\caption{(a) The minimum value of $U_t(c,X)$ as a function of $X$ for $F_0=0.8,x_c=1,\epsilon=0.01$ and $N=8$. (b) Parameter $c$ for the minimum value of $U_t(c,X)$ as a function of $X$ for $F_0=0.8,x_c=1,\epsilon=0.01$ and $N=8$. (c) Diffusion constant $D$ obtained by numerical simulation ($+$), estimated value by Eq.~(6) (dashed curve) and by Eq.~(2) (dotted curve)}
\label{f3}
\end{figure}
\begin{equation}
\frac{dX}{dt}=-F_0\sin(2\pi X)+\Xi(t),
\end{equation}
where $\Xi(t)=(1/N)\sum_{i=1}^N\xi_i(t)$ by the summation of Eq.~(1) for each $i$. The time correlation of $\Xi(t)$ satisfies $\langle \Xi(t)\Xi(t^{\prime})\rangle=(1/N^2)\sum_{j=1}^n 2T\delta(t-t^{\prime})=(2T/N)\delta(t-t^{\prime})$. 
The diffusion constant $D$ of $X(t)$ is evaluated as $D(T/N)$, using $D(T)$ in Eq.~(2) . This is a reason why $D$ is roughly approximated at $D(T/N)$ as shown in Fig.~2(a). However, the numerical values in Fig.~2(a) are slightly larger than $D(T/N)$ and the difference between the numerical results and the theoretical dashed curve  increases with $N$. Figure 2(b) shows the diffusion constant as a function of $T/N$ for three parameter values of $x_c$: $x_c=1,1.2$ and 1.4. The particle number is fixed to be $N=4$. As $x_c$ is increased, the diffusion constants approach the theoretical curve of $D(T/N)$. This is because more synchronous jump motion occurs as the jamming becomes stronger. 
Small deviation $\delta x_i=x_i-\{X(t)+i-(N+1)/2\}$ from the completely synchronous motion $x_i(t)=X(t)+i-(N+1)/2$ obeys 
\begin{equation}
\frac{d\delta x_i}{dt}=-2\pi F_0\cos(2\pi X(t))\delta x_i-F^{\prime}(1)(\delta x_{i-1}-2\delta x_i+\delta x_{i+1})+\xi_i(t)-\Xi(t),
\end{equation}
where $F^{\prime}(1)=-d^2U/dx^2=\epsilon(42\sigma^6-156\sigma^{12})=\epsilon(21x_c^6-39x_c^{12})$ are evaluated as -0.18, -2.85, and -20.53 respectively for $x_c=1,1.2$ and 1.4. As $x_c$ increases, the difference $\delta x_i-\delta x_{i-1}$  becomes smaller, and more synchronous motion occurs, because the interaction term expressed by $-F^{\prime}(1)(\delta x_{i-1}-2\delta x_i+\delta x_{i+1})$ works as more attractive force. As $N$ increases, the difference $\delta x_{N/2+1}-\delta x_1$ becomes larger and the synchronous motion becomes weaker owing to the fluctuations of long wavelength.

\begin{figure}[tbp]
\begin{center}
\includegraphics[height=5.cm]{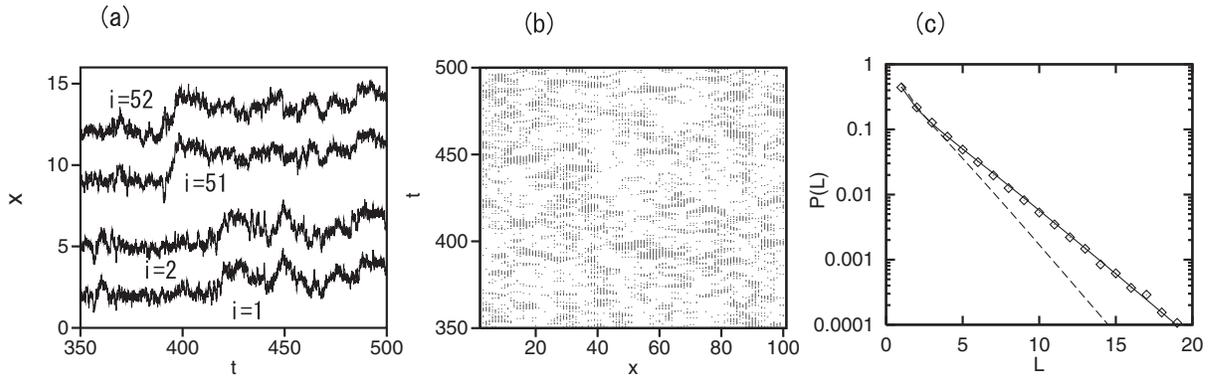}
\end{center}
\caption{(a) Time evolutions of $\tilde{x}_i$ for $i=1,2,51$ and 52 in a system of $x_c=1,F_0=0.8$ and $N=100$. (b) Spatio-temporal plot of the cooperative jump motions. (c) Numerically obtained probability distribution $P(L)$ (rhombi) of the size $L$ of the cooperative jump motions. An approximate probability distribution using Eq.~(7) is denoted by a dashed curve and its modification using Eq.~(8) with $\alpha=0.725$ is denoted by a solid curve.}
\label{f4}
\end{figure}
 The diffusion constant can be better evaluated by incorporating the deviation from the synchronous motion.   
We assume the deviation  $\delta x_i$ takes a form $\delta x_i=c\cos(2\pi i/N)$. Here, $c$ is a variational parameter which represents the amplitude of the sinusoidal deformation. The total energy of the configuration $x_i=i+X(t)-(N+1)/2+c\cos(2\pi i/N)$ is evaluated at 
\begin{equation}
U_t(c,X)=\sum_{i=1}^N \{U_0(x_i)+U(x_{i+1}-x_i)\}.
\end{equation}
We have calculated the minimum value of $U_t$ by changing $c$ for a fixed value of $X$. Figure 3(a) and 3(b) display the minimum value of $U_t$ and the parameter value of $c$ corresponding to the minimum of $U_t$ for $N=8$ and $x_c=1$. The synchronous state $c=0$ or $\delta x_i=0$ is unstable near $X=1/2$, because of the first term $-2\pi F_0\cos(2\pi X)\delta x_i$ in Eq.~(4). Too large deviation is unfavorable owing to the repulsive Lennard-Jones potential. 
As a result, a nonzero value of $c$ appears for $0.264<X<0.736$, and the peak amplitude of $U_t$ becomes smaller than $NU_0(1/2)$. That is, the deviation from the completely synchronized motion decreases effectively the peak potential energy, and the diffusion becomes easier. 
We found that the total fluctuation $\Delta x=\sum_{j=1}^N(x_{j+1}(t)-x_j(t))^2$ tends to increase rapidly when the jump motions occur as shown in Fig.~1(c). 
This observation is closely related to the above argument that the deviation from the completely synchronized motion becomes large when the particles overcome the potential peak at $X=1/2$. 
We have evaluated the diffusion constant as $D^{\prime}(T/N)$ using the modified potential $U_t$ as
\begin{equation}
D^{\prime}(T)=T\left\langle \exp\{U_t/(NT)\}\right\rangle^{-1}\left\langle \exp\{-U_t/(NT)\}\right\rangle^{-1}.
\end{equation}
Here, we have assumed that the energy $U_t/N$ is an effective periodic potential for the motion of $X$.     
Figure 3(c) displays the diffusion constant evaluated by Eq.~(6) with the dashed line. The dotted line is the one by Eq.~(2). The numerical results are denoted by the mark $\times$. The diffusion constant by Eq.~(6) is a better approximation. The diffusion constant becomes larger.  However, there is still a difference between the theoretical estimate and the numerical results. 
   
When the total number $N$ is much larger, the completely cooperative motion does not occur, but, the jump motion is locally synchronized. 
We show some numerical results for $N=100$. 
Figure 4(a) displays $\tilde{x}_1(t)+1,\tilde{x}_2(t)+3,\tilde{x}_{51}(t)-42$, and $\tilde{x}_{52}(t)-40$ for $N=100, \epsilon=0.01,x_c=1,F_0=0.8$ and $T=0.4$. The motions of $\tilde{x}_1$ and $\tilde{x}_2$ and the motions of $\tilde{x}_{51}$ and $\tilde{x}_{52}$ are sufficiently synchronized if large fluctuations are seen, but the motions of the two pairs are not synchronized. To characterize the jump motion, we have defined a quantity: $q(i,t)=\tilde{x}_i(t+\Delta t)-\tilde{x}_i(t)$.
If $q(i,t)$ is $O(1)$, a jump motion occurs for the $i$th particle during the interval between $t$ and $t+\Delta t$. Figure 4(b) shows a spatio-temporal plot of the jump events, where the spatio-temporal points satisfying $q(i,t)>0.8$ are plotted with dots for $\Delta t=5$. The jump motions occur cooperatively at sites in a horizontal line segment. This figure shows the dynamical heterogeneity in this system. The size of the cooperative jump motion or the number of particles which exhibit cooperative jump motions is evaluated with the length $L$ of the line segment satisfying $q(i,t)>0.8$ for fixed values of $t$. Figure 4(c) shows a semi-logarithmic plot of the size distribution  $P(L)$ for $N=100$ and $T=0.4$. The size distribution is approximately expressed with an exponential function.
The jump probability is proportional to the diffusion constant, and the diffusion constant has been roughly  evaluated as $D(T/N)=T/N\left\langle \exp\{NU_0/T\}\right\rangle^{-1}\left\langle \exp\{-NU_0/T\}\right\rangle^{-1}$ for a system including $N$ particles. Then, the cooperative jump probability including $L$ particles is expected to be proportional to $D(T/L)$ as a rough estimate. 
The dashed curve in Fig.~4(c) is 
\begin{equation}
P(L)=\frac{(T/L)\langle \exp(LU_0/T)\rangle^{-1}\langle \exp(-LU_0/T)\rangle^{-1}}{\sum_{L=1}^{\infty}(T/L)\langle \exp(LU_0/T)\rangle^{-1}\langle \exp (-LU_0/T)\rangle^{-1}}
\end{equation}
for $T=0.4$ and $U_0(x)=-F_0/(2\pi)\cos(2\pi x)$.  $P(L)$ decays exponentially, but the decay rate is larger than the numerical one.  As a better approximation, we take the effective decrease of the peak amplitude of the periodic potential by the deviation from the completely synchronized motion into consideration.  The effective periodic potential is assumed to be $\alpha U_0$ with $\alpha<1$ as a simple ansatz. 
Then, the size distribution is expressed as  
\begin{equation}
P(L)=\frac{(T/L)\langle \exp(L\alpha U_0/T)\rangle^{-1}\langle \exp(-L\alpha U_0/T)\rangle^{-1}}{\sum_{L=1}^{\infty}(T/L)\langle \exp(L\alpha U_0/T)\rangle^{-1}\langle \exp (-L\alpha U_0/T)\rangle^{-1}}.
\end{equation}
The solid curve in Fig.~4(c) denotes $P(L)$ calculated by Eq.~(8) with $\alpha=0.725$ for $T=0.4$. The solid curve is a good approximation, although the value of $\alpha$ is chosen as a fitting parameter.    
Donati et al. found cooperative string-like motions in the numerical simulations of supercooled liquids. They showed that the probability distribution of the string length obeys an exponential law~\cite{rf:6}. We think that our numerical result is closely related to their results. 

To summarize, we have proposed a very simple one-dimensional system which exhibits cooperative jump motions. We have found that the diffusion constant is roughly expressed as $D(T/N)$ where $N$ is the particle number. However, $D$ is slightly larger than the simplest one-particle approximation. We have tried a better approximation, taking the deviation from the completely synchronous motion into account. Finally, we have found that the probability distribution of the size of the cooperative jump motions obeys an exponential function of $L$. 
Our model is very simple, however, the theoretical analyses are still not satisfactory, which are left to future study.

\end{document}